\documentclass[a4paper,english,aps,prl,twocolumn,floatfix,showpacs,amsfonts,amssymb,superscriptaddress]{revtex4} 
\usepackage[latin1]{inputenc}
\usepackage{graphics}
\usepackage{epsfig}
\usepackage{amssymb}
\usepackage{amsmath}
\usepackage{overpic}
\usepackage{ulem}

\newcommand{\be}{\begin{equation}}
\newcommand{\ee}{\end{equation}}
\newcommand{\bea}{\begin{eqnarray}}
\newcommand{\eea}{\end{eqnarray}}

\def\a{\alpha}
\def\b{\beta}
\def\e{\varepsilon}

\def\l{\lambda}

\def\o{\omega}

\def\s{\sigma}

\def\G{\Gamma}

\def\L{\Lambda}

\def\ra{\rightarrow}

\def\pll{\parallel}

\def\pd{\partial}

\def\bk{{\bf k}}

\def\bq{{\bf q}}

\def\bQ{{\bf Q}}

\def\bJ{{\bf J}}

\def\bv{{\bf v}}

\def\nn{\nonumber}
\def\lb{\label}
\def\pref#1{(\ref{#1})}

\newcount\bozza \bozza=0
\ifnum\bozza=1
\newdimen\shift \shift=-2truecm
\def\lb#1{%
{\label{#1}\rlap{\kern\shift{$\scriptstyle#1$}}}}
\else\def\lb#1{\label{#1}} \fi

\begin{document}

\title{Unconventional Hall effect in pnictides from interband interactions}

\author{L.~Fanfarillo}
\affiliation{Institute for Complex Systems (ISC), CNR, U.O.S. Sapienza
  and Department of Physics,\\ Sapienza University of Rome, 
             P. le A. Moro 2, 00185 Rome, Italy}
\author{E.~Cappelluti}
\affiliation
{Instituto de Ciencia de Materiales de Madrid,
ICMM-CSIC, Cantoblanco, E-28049 Madrid, Spain}
\affiliation{Institute for Complex Systems (ISC), CNR, U.O.S. Sapienza
  and Department of Physics,\\ Sapienza University of Rome, 
             P. le A. Moro 2, 00185 Rome, Italy}
\author{C.Castellani}
\affiliation{Institute for Complex Systems (ISC), CNR, U.O.S. Sapienza
  and Department of Physics,\\ Sapienza University of Rome, 
             P. le A. Moro 2, 00185 Rome, Italy}
\author{L.~Benfatto}
\affiliation{Institute for Complex Systems (ISC), CNR, U.O.S. Sapienza
  and Department of Physics,\\ Sapienza University of Rome, 
             P. le A. Moro 2, 00185 Rome, Italy}

\date{\today}

\begin{abstract}
We calculate the Hall transport in a multiband systems with
a dominant interband interaction between carriers having electron and hole
character. We show that this situation gives rise to an unconventional
scenario, beyond the Boltzmann theory, where
the quasiparticle currents dressed by vertex corrections
acquire the character of the majority carriers. This leads to a larger
(positive or negative) Hall coefficient than what expected on the
basis of the carrier balance, with a marked temperature
dependence. Our results explain the puzzling 
measurements in pnictides and they provide a more general
framework for transport properties in multiband materials.
\end{abstract}


\maketitle

The discovery of superconductivity in iron-based superconductors has
triggered a renewed interest in the properties of interacting
multiband systems. Indeed, all the families of pnictides display
several hole (h) and electron (e) pockets at the Fermi level, as predicted by
density-functional-theory  calculations and confirmed by
Fermi-surface sensitive experiments, as de Haas van Alphen and
photoemission spectroscopy\cite{review_paglione}. A much more indirect probe of such a
multiband character comes from transport experiments, where the
contribution of carriers having h and e character is
unavoidably mixed. A typical example is provided by Hall-effect
measurements, for which a standard Boltzmann-like multicarrier picture\cite{kittel} would
give a Hall coefficient
\be 
\lb{rh0} 
R_H^0=\frac{1}{e}\frac{(n_h \mu^2_h-n_e\mu_e^2)}{(n_h\mu_h+n_e\mu_e)^2}, 
\ee 
where $n_\alpha$ ($\alpha=$e, h) is the density of the e- or h-
carrier type, and $\mu_\alpha=e^2\tau_\alpha/m_\alpha$ the
corresponding mobility, with $\tau_\alpha$ being the transport
scattering time and $m_\alpha$ the effective carrier mass in each
band. In compensated semimetals as pnictides, where $n_e\approx n_h$,
one would thus expect an almost vanishing Hall coefficient.
Striking enough, most pnictides 
\cite{rullier_hall,fang_PRB,rullier_Ru_prb10,matsuda_prb10,pallecchi_prb11,wen_prb11,borisenko_12}
 show a quite different scenario
with a very large absolute value of the Hall
coefficient $R_H$, and with a marked e/h character of the transport in
materials which are only slightly e/h doped.
In addition, a strong
temperature dependence of $R_H$ is also typically found, that
disappears only at very large doping away from half-filling.  Since
$\mu_\alpha\propto \tau_\alpha$, to account for these features within
the Boltzmann-like approach \pref{rh0} one needs thus to assume a
marked disparity between $\tau_e$ and $\tau_h$, with $\tau_e \gg
\tau_h$ in e-doped compounds and $\tau_h \gg \tau_e$ in the h-doped
ones. However, such a disparity has not been supported until now by
any explicit calculation. For example, the inclusion of
spin\cite{hirschfeld_prb11} or orbital\cite{kontani_prb12}
fluctuations within realistic models can account at most for a factor of
2 of anisotropy between the average quasiparticle lifetime on the
e and h pockets, not enough to explain neither the absolute
value of $R_H$  nor its $T$ dependence reported in
Refs. \cite{rullier_hall,fang_PRB,rullier_Ru_prb10,matsuda_prb10,pallecchi_prb11,wen_prb11,borisenko_12}.
In addition, the claim of two very different scattering rates in
pnictides obtained from optical probes, where the flat mid-infrared
optical conductivity is sometimes attributed to a very
broad Drude-like intraband contribution
\cite{dressel_prb10,homes_prb10,lucarelli_njop10}, has been questioned
by several authors on the basis of the presence in pnictides of
low-energy interband optical transitions
\cite{drechsler_prl08,homes_prb10,kuzmenko_epl10,benfatto_interband_prb11}.
A convincing framework to explain the unconventional properties of the
Hall transport in pnictides, and hence to elucidate the role of the
scattering mechanisms, is thus still lacking in these materials.

In this Letter we show that in a multiband system with predominant
interband interactions between carriers having opposite (e/h)
character the semiclassical picture of transport based on Eq.\
\pref{rh0} must be strongly revised.  By computing explicitly the
current vertex corrections due to the exchange of spin fluctuations
between h and e bands, we show that, in contrast to the
standard Fermi-liquid case, they cannot be simply recast in a
renormalization of the transport scattering time with respect to the
quasiparticle lifetime. Indeed, the spin fluctuations induce a mixing
of the electron and hole currents such that the renormalized current
in each band can even have opposite direction with respect to the bare
band velocity.  This mechanism explains the large value
of $|R_H|$ in slightly e/h doped compounds, and its temperature and doping
dependence, in good agreement with the experimental findings in the
non-magnetic state.

Let us introduce the minimal model which contains
the main ingredients responsible for the unconventional Hall transport in
pnictides. We consider a two-band model with two-dimensional parabolic e/h
bands centered at the $\Gamma$ and $M=\bQ=(\pi,\pi)$ points, with different
Fermi-surface (FS) areas: 
\bea
\lb{xih}
\xi^h_\bk=E^h_{max}-\frac{\bk^2}{2m_e}-\mu, \quad
\xi^e_{\bk}=-E^e_{min}+\frac{\bk^2}{2m_h}-\mu, 
\eea
where ${\bf k}$ is the reduced momentum
with respect to the $\Gamma$ or $M$ point,
for the h or e band, respectively. We take here
units $\hbar=c=a=1$ ($a$ being the
lattice spacing).  In the following we choose $\mu=0$, so that
$E^e_{min}$ and $E^h_{max}$ fix the Fermi wavevectors $k_F^{e,h}$
in each band, and we will
assume, without loss of generality, $m_e=m_h=m$. 
The general gauge-invariant expression for the longitudinal
and transverse conductivities for each band $\a$=e,h can be derived on
the basis of the Kubo formula for a weakly interacting system
\cite{fukuyama,kontani_prb99}. In particular we have that
\be
\lb{defsxx}
\s^\a_{xx} = e^2 \sum_{\bk} \Big(-\frac{\partial f}{\partial \xi }\Big)_{\xi^\a_\bk} 
 v^\a_x({\bk})  J_x^\a({\bk}) \frac{1}{\G^\a(\bk)}\simeq  \frac{e^2}{2} \frac{\bJ^\a_F\cdot  \bk^\a_F}{2 \pi \G^\a_F},
\ee
where the derivative of the Fermi function $f(x)=(1+e^{x/T})^{-1}$ has
been approximated at low temperature $T$
with a $\delta$-function, so
that only quantities at the FS appear in the final
expression. In Eq.\ \pref{defsxx} the vector $\bv^\a$
denotes the band velocity  in the $x$-$y$ plane for the band $\a$,
$\bJ^\a$ is the corresponding renormalized current and $\Gamma^\a$ is the inverse
quasiparticle lifetime, determined in general by electron-electron and
impurity scattering processes. Due to the symmetry of the
problem, $\bJ^\a$ is parallel to the reduced  moment $\bk$ in each
band $\a$, and 
$\Gamma^\a(\bk)$ and $\bJ^\a(\bk)$ depend only on $|\bk|$, so we
define their value at the FS as $\G^\a_F\equiv
\G^\a({k_F^\a)}$ and $\bJ^\a_F\equiv \bJ^\a(k_F^\a)$. 
The transverse $xy$
conductivity under a weak magnetic field $H$ along the $z$ axis
can be written as:
\be
\lb{defsxy} 
\frac{\s^\a_{xy}}{H} = - \frac{e^3}{4} \sum_\bk \Big(-\frac{\partial f}{\partial
  \xi }\Big)_{\xi^\a_\bk}
\frac{A^\a(\bk)}{(\G^\a(\bk))^2} \simeq \mp \frac{e^3}{8 \pi} \frac{(J^\a_F)^2}{(\G^\a_F)^2},
\ee
where $A^\a(\bk)=v^\a \left[\bJ^\a \times ({\bf e}^\a_{\parallel}
  \cdot \nabla) \bJ^\a\right] \cdot {\bf e}_z$. Here ${\bf e}_z$ is
the unit vector along the $z$ axis, while ${\bf e}^\a_\pll=({\bf e}_z
\times \bv^\a)/|\bv^\a|$ is tangential to the $\a$-th FS at $\bk$. For
a parabolic band we have thus ${\bf e}^\a_\pll\cdot \nabla=\pm
\pd_\theta/k$, where the plus/minus sign holds for an e/h band,
respectively, so that $A^\a=\pm (v^\a/k) (\bJ^\a \times \pd_\theta \bJ^\a)_z =\pm v^\a
(J^\a)^2/k$. As a consequence, the overall sign in Eq. \pref{defsxy}
is determined only by the sign of $\bv\cdot\bk$, which identifies the
e/h character of the band\cite{nota_kontani}. 
Once evaluated the longitudinal and transverse conductivity, the Hall
coefficient $R_H$ is given by
\be
\lb{RHmulti} 
R_H = \frac{\sum_{i} \sigma^i_{xy} }{(\sum_{i}
 \sigma^i_{xx})^2 H_z}.
\ee 

Eqs. (\ref{defsxx})-(\ref{RHmulti}) are quite general, since they
express the conductivities in terms of the bare Fermi velocity
$\bv^\a$ and the renormalized current $\bJ^\a$. What makes multiband
systems peculiar is the nature of vertex corrections that determine
the relation between $\bv^\a$ and $\bJ^\a$.  Using a standard
approach\cite{mahan} one can
establish between these two quantities a matricial relation
\be
\lb{gen} \bJ_F^\a =\Lambda_{\a\b} \bv_F^\b.
\ee 
The case
$\Lambda_{\a\b}=\delta_{\a\b}$ corresponds to the non-interacting
system where, by using the 2D relation
$n_\a=(\bk^\a_F)^2/2\pi$ and by identifying $1/\tau_\a=2\G^\a_F$, Eqs.\
\pref{defsxx}-\pref{defsxy} reduce to the standard results,
$\sigma^\a_{xx}=e^2 n_\a \tau_\a/m$ and $\sigma^\a_{xy}=\mp
\sigma^\a_{xx} \mu_\a H$, with $\mu_\a=e\tau_\a/m$, and the minus/plus
sign holds for the e/h band, respectively.  In the interacting case,
the strength of the diagonal and off-diagonal coefficients
$\Lambda_{\a\b}$ depends on the intraband or interband interactions,
respectively.  In conventional materials with predominance of intraband
scattering $\bJ_F^\a=\Lambda_{\a\a} \bv_F^\a$, so that the effect of
vertex corrections in Eq.\pref{defsxx} and \pref{defsxy} can be
reabsorbed in the definition of the transport scattering time
$\tau_\a=\Lambda_{\a\a}/2\Gamma^\a_F$, and the result \pref{rh0} still
holds, with renormalized mobilities. Things are however deeply
different in multiband systems with dominant interband interactions
connecting e and h sheets, as in pnictides.  In this case
the largest elements in $\Lambda_{\a\b}$ are off-diagonal, leading to
a mixing of the e and h characters and resulting in
unconventional features, like a possible vanishing current $\bJ^\a_F$.
In this situation, although the conductivities are still diagonal
in the band index $\a$ (as one can show following the general
derivation of Ref.\ \cite{fukuyama}), Eq. \pref{RHmulti} cannot be
reduced to the Boltzmann-like result \pref{rh0}.

To investigate in details this issue, in the following we compute
explicitly both $\G^\a_F$ and $\bJ^\a_F$ in the representative case of
pnictides, where the carriers in the h and e bands interact
via spin-fluctuations (SF) exchange\cite{review_paglione}.  According to neutron-scattering
experiments\cite{inosov_natphys10}, the SF spectrum can be
phenomenologically modeled with a standard marginal-Fermi-liquid
spectrum,
\be
\lb{mfl}
\chi(\bq-\bQ, \o) = \frac{\chi_Q}{1 + \xi_T^2(\bq-\bQ)^2 + i \o/\o_{sf}},
\ee
where $\chi_Q=\chi_0 \Theta/ (T+\Theta)$ is the strength of the SF,
$\o_{sf}=\o_0 (T+\Theta)/\Theta$ is their frequency scale and
$\xi_T=\xi_0 \sqrt{\Theta/(T+\Theta)}$ is the AF correlation length,
with $\Theta$ Curie-Weiss temperature. Since the SF are peaked around
the $\Gamma-M$ nesting vector $\bq=\bQ$, the interaction mediated by
such a collective mode will have a predominant interband
character. The crucial role of such interband retarded interaction has
been already demonstrated for the understanding of several
spectroscopic\cite{cappelluti_dhva,benfatto_cv,ummarino_prb09},
thermodynamic\cite{benfatto_cv,popovich_prl10} and
optical\cite{benfatto_sumrule} anomalies of pnictides.  However, to
compute current vertex corrections the explicit momentum dependent of
the bosonic spin spectrum \pref{mfl}, neglected so far in Refs.\
\cite{cappelluti_dhva,benfatto_cv,ummarino_prb09,popovich_prl10,benfatto_sumrule}, must be taken into account.
For the sake of simplicity we assume in the following that only
interband scattering is present, neglecting any intraband
coupling. The single-particle Green's function in each band is
computed as usual by means of the Dyson equation ${G^\a}^{-1}(\bk,
i\o_n) = i \o_n -\xi^\a_{\bk} -\Sigma^\a(\bk, i\o_n)$, where the
self-energy is given by
\be
\lb{defgh}
\Sigma^\a (\bk, \o_n) = g^2 T \sum_{\bq,l}  \chi({\bq},i\o_l) G^\b (\bk-\bq, \o_n-\o_l), 
\ee
where $\o_n,\o_l$ are fermionic and bosonic Matsubara frequencies,
respectively, and $g$ is the coupling to the bosonic mode
$\chi({\bq},i\o_l)$ of Eq.\ \pref{mfl}. In Eq.\ \pref{defgh} we 
accounted already for the nesting condition by considering only
interband terms, so that the most relevant fluctuations are around
$\bq=0$. As far as the vertex
corrections are concerned, following a standard derivation\cite{mahan,kontani_prb99},
the current at $\o=0$ is computed as 
%
\bea
\bJ^\a(\bk)
&=&
\bv^\a(\bk)+
g^2  \sum_\bq \int_{-\infty}^\infty \frac{d \e}{2\pi} 
F(\e) {G^\b}^R (\bk+\bq,\e)  \nn \\
\lb{djint}
&\quad&  \times {G^\b}^A (\bk+\bq,\e)  \text{Im} \chi^R(\bq, \e) \bJ^\b({\bk+\bq}).
\eea
where $F(\e)=\coth (\e/2 T) - \tanh (\e/2 T)$ and $G^{R,A}$ is the
retarded/advanced Green's function. 
\begin{figure}[t]
\begin{center}
\includegraphics[clip=true,scale=.3,angle=0,clip=true]{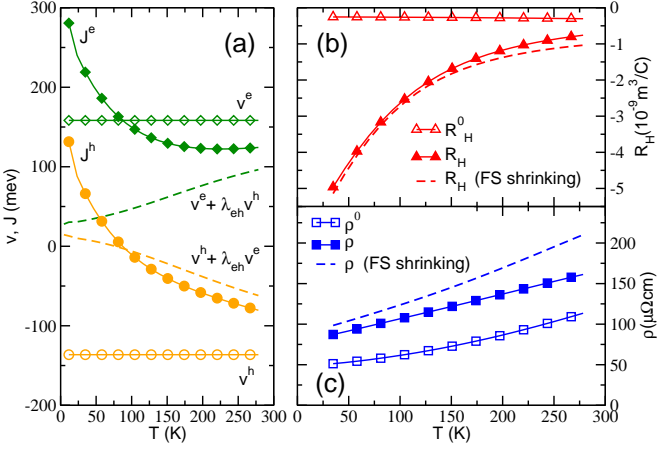}
\end{center}
\caption{(a) $T$ dependence of the renormalized currents
  (filled symbols), as
  compared to the bare velocities (empty symbols).  We
  also show (dashed lines) the numerators of Eqs.\ \pref{jh}-\pref{je}, that fix the
  overall sign of the currents. (b) $T$ dependence of the Hall
  coefficient $R_H$ compared to the Boltzmann result 
  \pref{rh0} $R_H^0$, computed with $1/\tau^\a=2\G_F^\a$. The units
  are fixed by the two-dimensional results divided by the
  interlayer distance d = 6.5 A.  Dashed line: $R_H$ obtained
  including also the effect of the FS shrinking. (c) Longitudinal
  resistivity as a function of $T$ compared to $\rho^0=\left( \sum_\a e^2n_\a \tau^\a/m\right)^{-1}$ 
in the Boltzmann approximation  with $1/\tau^\a=2\G_F^\a$.} 
\label{fig-curr}
\end{figure}
To compute the conductivities \pref{defsxx}-\pref{defsxy} we need the
quasiparticle scattering rates
$\Gamma_F^\a$ and the dressed currents $\bJ_F^\a$ at the
Fermi wavevectors $\bk^\a_F$. By accounting for the most relevant contributions to
the integrals \pref{defgh}-\pref{djint} one can obtain an approximated
semi-analytical expression for all these quantities\cite{suppl}. The imaginary part of
the self-energy is thus given by:
\be
\lb{gamma}
\G^\a_F = \frac{g^2 \o_{sf} \, \chi_Q }{2} \sum_\bq  
F(\xi^\b_{\bk_F^\a+\bq}) 
\frac{ \xi^\b_{\bk_F^\a+\bq}}{\o_\bq^2 + (\xi^\b_{\bk_F^\a+\bq})^2} ,
\ee
where $\o_\bq =\o_{sf}(1 + \xi_T^2\bq^2)$. At the same time by
introducing the velocity and current projection along $\bk$, i.e.  
$J^\a\equiv \bJ^\a_F\cdot \hat\bk$ and $v^\a\equiv
\bv^\a_F\cdot \hat\bk$, so that $v^h<0$ and $v^e>0$, 
one finds\cite{suppl} for the renormalized currents the expression
\pref{gen} above
\bea
\lb{jh}
J^h&=& (1-\l_{he} \l_{eh})^{-1} (v^h+\l_{he} \ v^e)  \\
\lb{je}
J^e&=& (1-\l_{he} \l_{eh})^{-1} (v^e+\l_{eh} \ v^h) 
\eea
where the matrix $\L_{\a\b}$ of Eq.\ \pref{gen} has been expressed in terms of the
temperature-dependent coefficients
\be
\lambda_{\a\b}=\frac{g^2\o_{sf} \, \chi_Q}{2 \G^\b_F}
\sum_\bq F(\xi^\b_{\bk_F^\a+\bq})\frac{\xi^\b_{\bk^\a+\bq}}{\o_\bq^2 + (\xi^\b_{\bk^\a+\bq})^2}
\frac{\bk^\a_F + \bq \cos \theta_q}{|\bk_F^\a+\bq|}.
\lb{alpha}
\ee
By close inspection of Eqs.\pref{gamma} and \pref{alpha} we see that
$\l_{\a\b}$ increases as the nesting condition is approached,
since the largest contribution to the integrals comes from vectors
around $\bar \bq\sim \bk_F^\a-\bk_F^\b$. Moreover, at high $T$ where
$\xi_T\simeq 0$, so that $\o_\bq$ is independent on $\bq$, 
$\lambda_{\a\b}$ vanishes and $J^\a=v^\a$. Conversely, at low $T$, $\xi_T$
increases and only the value
$\theta_{\bar \bq}=0$ contributes to the integral \pref{alpha},
leading to $\l_{\a\b}\ra 1$ and large prefactors in Eqs.\
\pref{jh}-\pref{je}.  

To elucidate the effect on the transport of such a scattering
mechanism connecting e- and h-like bands we will consider a set of
parameters appropriate for electron-doped
Ba(Fe$_{1-x}$Co$_x$)$_2$As$_2$. In particular we take $1/2m=70$ meV,
and we choose $E^e_{min}=90$ meV and $E_{max}=66$ meV (i.e
$k_{F}^h=0.30 \pi/a$ and $k_{F}^e=0.37 \pi/a$) to reproduce the data at
$7\%$ doping, where long-range AF order is no more present and our
model applies. For the SF we use
\cite{inosov_natphys10} $\o_0=15$ meV, $\Theta=90$ K, $\xi_0/a=3.6$
and $g^2\chi_Q=0.8$ eV. Since the low-energy description \pref{mfl} is
not expected to hold any more around a scale of the order of the room
temperature , we rescaled $\xi_T=\xi_0
\sqrt{\Theta/(T+\Theta)}\exp(-T/T_{cut})$ to account for a fast decay
of AF correlations above $T_{cut}\simeq 300$ K. Finally, to mimic the
residual scattering by impurities at $T=0$ we added a constant
(isotropic) scattering rate $\G_0^\a=4$ meV. The resulting currents for
each band as a function of temperature, as evaluated from
Eqs. (\ref{gamma})-(\ref{alpha}), are shown in Fig.\ \ref{fig-curr}a,
along with the bare Fermi velocities. Two relevant features
emerge. First, we find a strong temperature dependence of both $J^e$
and $J^h$, which deviate significantly from their bare values with
lowering temperature, due to the increasing scattering from
SF. Second, in the e-doped case considered here, where $|v_e|>|v_h|$,
the dominance of $\l_{he}v_e$ with respect to $v_h$ in Eq. (\ref{jh})
is reflected in a change of sign of $J^h$ at low
$T$. These features have a striking effect on the Hall transport,
shown in Fig.\ \ref{fig-curr}b, along with the Boltzmann result
\pref{rh0} computed without vertex corrections. As expected, $R_H^0$
is small and weakly $T$ dependent, as due to the almost perfect
cancellation of the contributions from the h- and e-like Fermi
sheets. On the contrary $R_H$ has a strong temperature dependence
and it can attain
a large negative value at low $T$, where the e-like renormalized
current $|J^e|\gg |J_h|$ dominates the transverse conductivity
\pref{defsxy}.  At the same time, the effects of the vertex
renormalization are less qualitatively relevant on the longitudinal
resistivity with respect to the Boltzmann result, as shown in Fig.\
\ref{fig-curr}c. Indeed, the dependence of $\s^\a_{xx}$ in Eq.\
\pref{defsxx} on the sign of the renormalized current leads to a
compensation between the vertex corrections in the h and e bands.  Our
results provide thus a consistent picture for both longitudinal and
transverse transport, in good agreement with the experimental
findings \cite{rullier_hall,fang_PRB}.  For the sake of completeness
we show in Fig.\ \ref{fig-curr}b,c also the effect of the weakly
temperature-dependent FS shrinking
arising from the real part of the self-energy \pref{defgh}\cite{cappelluti_dhva,
  benfatto_sumrule,suppl}. As one can see, this is irrelevant on the Hall transport, while it
contributes in part to the temperature dependence of the longitudinal
conductivity, as discussed in Ref.\ \cite{benfatto_sumrule}.

\begin{figure}[t]
\hspace{-1.5cm}
\begin{minipage}[h]{0.35\linewidth}
\centering
\vspace{0.5cm}
\includegraphics[scale=0.27,angle=-90, clip=]{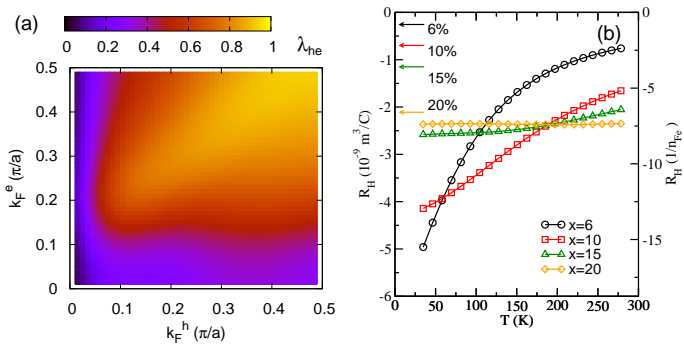}
\end{minipage}
\hspace{1.4cm}
\begin{minipage}[h]{0.35\linewidth}
\centering
\includegraphics[scale=0.25,clip=]{fig2b.eps}
\end{minipage}
\caption{(Color online) (a) Dependence of the vertex-correction
  coefficient $\l_{he}$ at $T=100$ K on the Fermi-wavevectors in the
  two bands. Notice that $\l_{he}$ decreases as one moves away from
  the nesting line $k_F^e=k_F^h$. (b) $R_H$ as a function of the Co concentration $x$ (which
  adds $x$ electrons per Fe atom) The arrows indicate the corresponding values of
  the $T=0$ Boltzmann result \pref{rh0}, which has a negligible $T$
  dependence on this scale (see Fig.\ \ref{fig-curr}b).  On the
  right axes $R_H$ is expressed in units of the inverse number of
  carriers per $Fe$ atom, defined as $n_{Fe}=0.32 \times
  10^{-9}/|R_H[m^3/C]|$.}
\label{fig-map}
\end{figure}
It is also interesting to address the effects of
doping. Indeed, as we mentioned above and as we show in Fig.\
\ref{fig-map}a, the absolute value of the $\lambda_{\a\b}$
coefficients decreases as the Fermi wavevectors move away from the
nesting condition $k_F^e=k_F^h$, realized at half-filling.  We expect
thus that the effect of the vertex corrections on the Hall transport
will be less relevant by further increasing the Co concentration
$x$. We investigate this issue by making a rigid-band shift of the
chemical potential with doping, without changing for simplicity the
microscopical parameters of the SF spectrum.  The resulting Hall
coefficient is reported in Fig.\ \ref{fig-map}b, where we also mark
with arrows the corresponding Boltzmann value $R_H^0$ at each
doping. As one can see, the low-$T$ enhancement of $R_H$ induced by SF
decreases with increasing doping, and for $x=20\%$ $R_H$ almost
coincides with $R_H^0\simeq 1/e x$, as found
experimentally\cite{rullier_hall}. 
The trend shown in Fig.\ \ref{fig-map}b,
where SF spectrum is kept constant, is already in good agreement
with the experiments. Nonetheless, one could also expect a decrease of
the AF correlation length $\xi_T$ with doping, leading to a faster
suppression of vertex corrections. This effect could explain the
results in isovalent-substituted systems, as for example
BaFe$_2$(As,P/Ru)$_2$\cite{rullier_Ru_prb10,matsuda_prb10} or
La(Fe,Ru)AsO\cite{pallecchi_prb11}, where the change of magnitude (or
even of the sign) of the Hall coefficient should be attributed to a
weakening of AF correlations, since no significant change on the FS
pockets seems to occurs\cite{kaminski_prl11}.  Finally, we notice that
for a hole-doped system the overall temperature and doping dependence
of $R_H$ would be exactly the specular one: indeed, when the system is
doped with holes, one has in general $k_F^h>k_F^e$, so that $|J^h|\gg
|J^e|$ at low $T$ and the transverse conductivity will have a
predominant hole-like character, in agreement with the
experiments\cite{wen_prb11, borisenko_12}. Thus, the same mechanism
accounts for the unusual Hall effect measured in pnictides both in
electron and hole-doped compounds.

In conclusion we analyzed the Hall effect in a multiband model where
carriers interact via the exchange of SF. We showed that when
interactions have a predominant interband character and connect
carriers of opposite e/h nature, the currents renormalized by vertex
corrections are dominated by the character of the majority
carriers. By evaluating this effect within a simplified two-band model
and a phenomenological description of SF we were able to reproduce the
main puzzling features observed experimentally in pnictides, namely a
strong temperature dependence of $R_H$ with a large absolute value at
low $T$ for weak (e or h) doping, and a more ordinary Boltzmann-like
behavior at higher doping. We notice that the mechanism discussed here
is quite general and robust: thus, an analysis based on a full
self-consistent approach for the SF\cite{kontani_prb99} within
microscopic multiband models\cite{hirschfeld_prb11,kontani_prb12} is
expected to add only quantitative refinements to the present
results. An open question is instead the role of vertex corrections
across the antiferromagnetic transition, where the Hall coefficient
has been found experimentally to show an even larger $T$
dependence. Even though this issue is beyond the scope of the present
manuscript, it certainly deserves further investigation to complete our
theoretical understanding of transport properties in pnictides.

E.C. acknowledges support from the European FP7 Marie
Curie project PIEF-GA-2009-251904.


\begin{thebibliography}{99}

\bibitem{review_paglione}
For a review see e.g. J. Paglione and R. L. Green, Nat. Phys. 6, 645 (2010).

\bibitem{kittel}
C.~Kittel, {\it Introduction to Solid State Physics} (Wiley, New York, 2005) 

\bibitem{rullier_hall}
F.~Rullier-Albenque, {\it et al.}
Phys. Rev. Lett. {\bf 103}, 057001 (2009); 

\bibitem{fang_PRB}
L.~Fang, {\it et al.}
Phys. Rev. B {\bf 80}, 140508(R) (2009). 

\bibitem{rullier_Ru_prb10}
F.~Rullier-Albenque,  {\it et al.}
\prb {\bf 81}, 224503 (2010) 

\bibitem{matsuda_prb10}
S.~Kasahara, {\it et al.}
\prb {\bf 81}, 184519 (2010) 

\bibitem{pallecchi_prb11}
I.~Pallecchi, {\it et al.}
\prb {\bf 84}, 134524 (2011) 

\bibitem{wen_prb11}
B.~Shen, {\it et. al.}
\prb {\bf 84}, 184512 (2011) 

\bibitem{borisenko_12}
S.~Aswartham, {et al.} 
arXiv:1203.0143 (2012) 

\bibitem{hirschfeld_prb11}
A.F.~Kemper, {\it et al.}
Phys. Rev. B {\bf 83}, 184516 (2011). 

\bibitem{kontani_prb12}
S.~Onari and H.~Kontani, Phys. Rev. B {\bf 85}, 134507 (2012).

\bibitem{dressel_prb10}
N.~Barisic, {\it et al.}
Phys. Rev. B {\bf 82}, 054518 (2010).

\bibitem{homes_prb10}
J.J.~Tu, {\it et al.}
Phys. Rev. B {\bf 82}, 174509 (2010). 

\bibitem{lucarelli_njop10}
A.~Lucarelli et al., New J. of Phys. {\bf 12}, 073036 (2010)

\bibitem{kuzmenko_epl10}
E.~van Heumen, {\it et al.}
Europhys. Lett. {\bf 90}, 37005 (2010). 

\bibitem{drechsler_prl08}
S.L.~Drechsler, {\it et al.}
Phys. Rev. Lett. {\bf 101}, 257004 (2008)

\bibitem{benfatto_interband_prb11}
L.~Benfatto, E.~Cappelluti, L.~Ortenzi, and L.~Boeri,
Phys. Rev. B {\bf 83}, 224514 (2011)


\bibitem{kontani_prb99}
H.~Kontani, K.~Kanki and K.~Ueda,
Phys. Rev. B {\bf 59}, 14723 (1999). 

\bibitem{fukuyama}
H.~Fukuyamha, H.~Ebisawa and Y.~Wada,
Prog. Theor. Phys. {\bf 42}, 494 (1969)

\bibitem{nota_kontani} 
Notice that we will not consider here possible anomalies in the Hall
effect due the band anisotropy, as discussed for example in
Ref. \cite{kontani_prb99} within the context of cuprates,
where the band dispersion deviates significantly from a parabolic
one and $\bJ$ and $\bv$ in general are not  parallel.

\bibitem{mahan}
G.~D.~Mahan, {\it Many-Particle Physics}, Plenum Press, 1981.

\bibitem{inosov_natphys10}
D.S.~Inosov,  {\it et al.}
Nature Phys. {\bf 6}, 178 (2010). 

\bibitem{cappelluti_dhva}
L.~Ortenzi, {\it et al.}
Phys. Rev. Lett. {\bf 103}, 046404 (2009).

\bibitem{benfatto_cv}
L.~Benfatto, E.~Cappelluti and C.~Castellani, 
Phys. Rev. B {\bf 80}, 214522 (2009). 

\bibitem{ummarino_prb09}
G.A.~Ummarino, {\it et al.}
Phys. Rev. B {\bf 80}, 172503 (2009) 

\bibitem{popovich_prl10}
P.~Popovich, {\it et al.}
Phys. Rev. Lett. {\bf 105}, 027003 (2010) 

\bibitem{benfatto_sumrule}
L.~Benfatto and E.~Cappelluti, 
Phys. Rev. B {\bf 83}, 104516 (2011).


\bibitem{suppl}
Additional details are reported in Supplementary Information.

\bibitem{kaminski_prl11}
R. S. Dhaka, {\it et al.}
Phys. Rev. Lett. {\bf 107}, 267002 (2011) 

\end{thebibliography}
\end{document}